\pdfoutput=1
\documentclass[aip,jcp,reprint]{revtex4-2}
\usepackage{physics}
\usepackage{amsmath,amsfonts,amssymb}
\usepackage{bm}
\usepackage{comment}
\usepackage{graphicx, float}

\usepackage{dcolumn}%
\usepackage[T1]{fontenc}
\usepackage{mathptmx}
\usepackage{etoolbox}

\usepackage[acronym]{glossaries}
\usepackage[dvipsnames,table]{xcolor} %
\usepackage[version=3]{mhchem}
\usepackage{natmove} % Not needed for achemso but necessary for revtex4-1

\newacronym[longplural={degrees of freedom}, firstplural={degrees of
  freedom (DOF)}, plural={DOF}]{DOF}{DOF}{degree of freedom} %
\newacronym[longplural={equations of motion}, firstplural={equations
  of motion (EOM)}, plural={EOM}]{EOM}{EOM}{equation of motion} %
\newacronym{TDSE}{TDSE}{time-dependent Schr\"odinger equation}
\newacronym{TIDSE}{TIDSE}{time-independent Schr\"odinger equation}
\newacronym{TDVP}{TDVP}{time-dependent variational principle} %
\newacronym{DFVP}{DFVP}{Dirac--Frenkel variational principle} %
\newacronym{MCTDH}{MCTDH}{multiconfguration time-dependent Hartree}

\newacronym{vMCG}{vMCG}{variational multi-configurational Gaussian}
\newacronym{FMS}{FMS}{full multiple spawning} %
\newacronym{G-MCTDH}{G-MCTDH}{Gaussian-based multi-configuration time-dependent Hartree} %
\newacronym{MAE}{MAE}{mean absolute error} %
\newacronym{pMAER}{\%MAER}{percentage of mean absolute error reduction}

\newacronym{GWP}{GWP}{Gaussian wavepacket} %
\newacronym{TGWP}{TGWP}{thawed Gaussian wavepacket} %
\newacronym{GHA}{GHA}{global harmonic approximation} %
\newacronym{LHA}{LHA}{local harmonic approximation} %
\newacronym{ZPE}{ZPE}{zero-point energy} %
\newacronym{VM}{VM}{vibrational mode} %
\newacronym{CVM}{CVM}{curvilinear vibrational mode}

\newacronym{ML}{ML}{machine learning} %
\newacronym{ML-PES}{ML-PES}{machine-learned potential energy surface} %
\newacronym{KRR}{KRR}{kernel ridge regression} %
\newacronym{GPR}{GPR}{Gaussian process regression} %

\newacronym{OLED}{OLED}{organic light-emitting diode}
\newacronym{NISQ}{NISQ}{noisy intermediate-scale quantum}

\newacronym{JW}{JW}{Jordan--Wigner} %
\newacronym{BK}{BK}{Bravyi--Kitaev} %

\newacronym{QPE}{QPE}{quantum phase estimation} %
\newacronym{VQE}{VQE}{variational quantum eigensolver} %
\newacronym{QMF}{QMF}{qubit mean-field} %
\newacronym{QCC}{QCC}{qubit coupled cluster} %
\newacronym{iQCC}{iQCC}{iterative qubit coupled cluster} %
\newacronym{PQA}{PQA}{parametrized quantum annealing} %
\newacronym{DIS}{DIS}{direct interaction set} %
\newacronym[longplural={involutary linear combinations of
anti-commuting Paulis}, firstplural={involutory linear combinations of
anti-commuting Paulis (ILCAP)}, plural={ILCAP}]{ILCAP}{ILCAP}{involutory linear combination of anti-commuting Paulis} %

\newacronym{CAS}{CAS}{complete active space} %
\newacronym{PES}{PES}{potential energy surface} %
\newacronym{PEC}{PEC}{potential energy curve} %
\newacronym{AO}{AO}{atomic orbital} %
\newacronym{MO}{MO}{molecular orbital} %

\newacronym{CI}{CI}{configuration interaction} %
\newacronym{FCI}{FCI}{full configuration interaction} %
\newacronym{CASCI}{CASCI}{complete active space configuration interaction} %
\newacronym{MCSCF}{MCSCF}{multiconfigurational self-consistent
  field} %
\newacronym{CASSCF}{CASSCF}{complete active space self-consistent
  field} %
\newacronym{CC}{CC}{coupled cluster} %
\newacronym{UCC}{UCC}{unitary coupled cluster} %
\newacronym{UCCSD}{UCCSD}{unitary coupled cluster singles and
  doubles} %
\newacronym{CCSD}{CCSD}{coupled-cluster singles and doubles} %
\newacronym{CCSD-T}{CCSD(T)}{coupled-cluster singles and doubles and
  non-iterative triples} %
\newacronym{RHF}{RHF}{restricted Hartree--Fock} %
\newacronym{CIS}{CIS}{configuration-interaction singles} %
\newacronym{ROHF}{ROHF}{restricted open-shell Hartree--Fock} %
\newacronym{UHF}{UHF}{unrestricted Hartree--Fock} %
\newacronym{DMRG}{DMRG}{density-matrix renormalization group} %
\newacronym{DFT}{DFT}{density-functional theory} %
\newacronym{TDDFT}{TDDFT}{time-dependent density-functional theory} %
\newacronym{ENPT}{ENPT}{Epstein-Nesbet perturbation theory} %
\newacronym{MP}{MP}{M{\o}ller--Plesset perturbation theory} %
\newacronym{MP2}{MP2}{second-order M{\o}ller--Plesset perturbation theory} %
\newacronym{MRMP2}{MRMP2}{second-order multi-reference M{\o}ller--Plesset perturbation theory} %

\newacronym{SQP}{SQP}{sequential quadratic programming} %
\newacronym{MMA}{MMA}{method of moving asymptotes} %

\begin{document}

\title{Thawed Gaussian wavepacket dynamics with $\Delta$-machine
  learned potentials}

\author{Rami Gherib}
\email{rami.gherib@otilumionics.com}

\author{Ilya G. Ryabinkin}
\email{ilya.ryabinkin@otilumionics.com}

\author{Scott N. Genin}
\email{scott.genin@otilumionics.com}

\affiliation{OTI Lumionics Inc., 3415 American Drive Unit 1, Mississauga, Ontario L4V\,1T4, Canada}

\date{\today}

\begin{abstract}
  A method for performing variable-width (thawed) \gls{GWP}
  variational dynamics on machine-learned potentials is presented.
  Instead of fitting the \gls{PES}, the anharmonic correction to the
  \gls{GHA} is fitted using \acrlong{KRR}---this is a $\Delta$-machine
  learning approach. The training set consists of energy differences
  between \emph{ab initio} electronic energies and values given by the
  \gls{GHA}. The learned potential is subsequently used to propagate a
  single thawed \gls{GWP} using the \acrlong{TDVP} to compute the
  autocorrelation function, which provides direct access to vibronic
  spectra via its Fourier transform. We applied the developed method
  to simulate the photoelectron spectrum of ammonia and found
  excellent agreement between theoretical and experimental spectra. We
  show that fitting the anharmonic corrections requires a smaller
  training set as compared to fitting total electronic energies. We
  also demonstrate that our approach allows to reduce the
  dimensionality of the nuclear space used to scan the \gls{PES} when
  constructing the training set. Thus, only the degrees of freedom
  associated with large amplitude motions need to be treated with
  $\Delta$-machine learning, which paves a way for reliable
  simulations of vibronic spectra of large floppy molecules.
\end{abstract}

\maketitle

\glsresetall

\section{Introduction}

Computational spectroscopic methods play an important role in
\textit{in silico} predictions of novel \acrlong{OLED} materials. One of
the essential characteristics of such materials is their
vibrationally-resolved electronic (vibronic) spectra. The vibronic
spectrum of a given molecule can be obtained through the Fourier
transform of its autocorrelation
function~\cite{heller1981semiclassical, tannor2007introduction,
  niu2010theory, peng2010vibration, baiardi2013general}. In turn,
constructing an autocorrelation function requires solving the
time-dependent Schr\"{o}dinger equation. For sizable molecules, this
task is known to be difficult. Numerically exact methods such as the
split-operator~\cite{kosloff1988time} or the
\gls{MCTDH}\cite{meyer2018mctdh, beck2000multiconfiguration} methods
scale exponentially with the number of nuclei and cannot be applied to
industrially relevant molecules.

Fortunately, exact solutions are not always needed and approximate
schemes are often sufficient. In particular, \gls{GWP}
methods~\cite{sawada1986gaussian, worth2004novel, curchod2018ab,
  worth2020gaussian, joubert2022variational, jl2023family} are
well-studied practical alternatives. This family of methods represents
the nuclear wavefunction as a linear combination of multidimensional
Gaussians. Different schemes are used to evolve the \gls{GWP}
parameters and expansion coefficients. For example, the
\gls{vMCG}~\cite{richings2015quantum} approach evolves Gaussian
positions and expansion coefficients according to the \gls{TDVP}.
Others, like the \acrlong{FMS} method~\cite{ben2000multiple}, evolve
the positions and momenta of Gaussian functions according to classical
mechanics.

A subset of \gls{GWP} methods represents the entire wavefunction as a
single multidimensional Gaussian. This is in part motivated by a
standard result in quantum mechanics, which shows that for the case
of a harmonic potential, a \gls{GWP} maintains that form
during its evolution. Even in cases where the potential is anharmonic,
a single GWP can often be a reasonable approximation, especially when
the \gls{PES} varies slowly on the scale of its
width~\cite{wehrle2015fly, begusic2022applicability}, and when the
relevant timescale is short. Arguably, one of the most popular single
\gls{GWP} schemes is the semiclassical approach proposed by
Heller~\cite{heller1975time} in which the \gls{GWP} center follows the
classical trajectory.

\Gls{GWP} methods that are based on the \gls{TDVP} require the
evaluation of potential energy matrix elements, which are
multidimensional integrals involving electronic potentials. This
necessitates \glspl{PES} to be represented in compact and
computationally efficient form, mitigating or eliminating the curse of
dimensionality, while at the same time ensuring that multidimensional
integrals can be evaluated efficiently. One approach, often referred
to as the \gls{LHA}, consists of Taylor expanding and truncating of
the molecular \gls{PES} to second order around the center of each
\gls{GWP}. The \gls{LHA} has the advantage of exploiting the localized
nature of \glspl{GWP}, but suffers from the drawbacks of requiring
computational costly Hessian (second-order derivatives) matrix
calculations at each time step. Additionally, the potential energy
matrix elements are correct only to the second
order~\cite{frankcombe2010converged, polyak2019direct}.

To address these limitations, some studies have proposed to fit
\glspl{PES} using \gls{ML} and to employ machine-learned potentials in
\gls{GWP} dynamics. For instance, \citet{alborzpour2016efficient}
simulated nuclear dynamics with a wavefunction ansatz consisting of a
linear combination of Gaussians whose centers followed classical
trajectories on \gls{PES} fitted by \gls{GPR}.
\Citet{richings2017direct} used on-the-fly fitting of \glspl{PES} by
\gls{GPR} in \gls{MCTDH} simulations. \Citet{polyak2019direct} also
employed \glspl{PES} fitted with \gls{GPR} in \gls{vMCG} simulations.
Finally, \citet{koch2019two} performed two-layer \acrlong{G-MCTDH}
dynamics on multiplicative neural network potentials utilizing
exponential transfer functions.

A significant portion of the computational efforts of producing
\glspl{ML-PES} stems from the electronic structure calculations needed
to build the training set. Obviously, one way to reduce the
computational burden is to reduce the size of the training set. In
this study, we show how this can be accomplished in a relatively
simple way. We demonstrate that by fitting \gls{ML} models on
anharmonic corrections, rather than on the \glspl{PES} themselves, we
can significantly reduce the number of single-point electronic
structure calculations needed to fit \glspl{ML-PES}. This approach is
a type of $\Delta$-machine learning method~\cite{ramakrishnan2015big,
  ramakrishnan2015electronic}, which has been utilized in a previous
study to correct energies predicted by \gls{CCSD}, taken as reference,
to bring them closer to energies produced by \gls{CCSD-T}
\cite{ruth2022machine}. In our approach we choose the reference method
to be the \gls{GHA}, and the higher-level of theory to be the energies
given by DFT calculations. The \gls{GHA} represents the entire
\gls{PES} as a paraboloid constructed from the truncated Taylor
expansion at some fixed nuclear configuration. Unlike the \gls{LHA},
the \gls{GHA} requires gradient and Hessian calculations to be
performed only once throughout all the dynamics.

This study uses \gls{KRR}~\cite{vovk2013kernel, unke2017toolkit,
  dral2017structure, hu2018inclusion, westermayr2020neural,
  dral2020quantum, pinheiro2021choosing} to construct \glspl{ML-PES}.
It is a non-parametric \gls{ML} technique capable of fitting
non-linear multidimensional functions and contains a regularization
parameter to prevent overfitting. Compared to some other \gls{ML}
techniques, such as artificial neural networks, \gls{KRR} is typically
easier to implement and generally performs better, especially when one
is limited to a relatively small training
set~\cite{pinheiro2021choosing}. An additional advantage is that
\gls{KRR} with a Gaussian kernel can be quite naturally integrated
with \gls{GWP} dynamics.
% Numerically, \gls{KRR} is equivalent to
% \gls{GPR}, which has already been used in the context of \gls{GWP}
% dynamics~\cite{alborzpour2016efficient, polyak2019direct}.

We investigate the $\Delta$-\gls{KRR} scheme of fitting \glspl{ML-PES}
in conjunction with the variational dynamics of a single thawed
multidimensional Gaussian. As numerical validation, we simulate the
photoelectron spectra of ammonia (\ce{NH3}), which is commonly used as
a prototype of floppy molecules. We show how the $\Delta$-\gls{KRR}
scheme not only helps to reduce the size of the training set but also
allows to reduce the dimensionality of the nuclear subspace that is
used to build the training set.

The rest of the manuscript is organized as follows. First, we present
the working equations needed to perform time-dependent variational
dynamics of a single \gls{TGWP}, while assuming a particular
decomposition of the \gls{PES} separating harmonic and anharmonic
contributions. Next, we present the mathematical background detailing
how $\Delta$-\gls{KRR}-derived \glspl{PES} can be incorporated into
our \gls{GWP} dynamics. We also describe the methodology we used to
construct training sets suitable for molecular dynamics along
curvilinear coordinates. Finally, we validate the approach on ammonia,
a molecule on which the \gls{GHA} performs poorly. Mathematical
symbols representing arrays are written in bold. To improve the
readability of the mathematical expressions, two-dimensional arrays
are denoted with a check mark ($~ \check{ } ~$) to distinguish them
from one-dimensional arrays. Atomic units are used throughout.

\section{Theory}

\subsection{Time-dependent vibronic spectroscopy}

In this study, we consider dipole-allowed optical transitions between
only two relevant electronic states. The initial and final electronic
states are denoted as $\ket{\Phi_{\text{I}}}$ and
$\ket{\Phi_{\text{F}}}$, respectively. We assume that the system is
initially in the vibrational ground state $\chi_{{0}}(\bm{R})$ of
$\ket{\Phi_{\text{I}}}$. The nuclear wavepacket $\chi(\bm{R}, t)$
evolves on the single final state \gls{PES} represented by a function
$V_{\text{F}}(\bm{R})$; $\chi(\bm{R},0) = \chi_{0}(\bm{R})$.
Absorption spectra $I_\text{abs}(\omega)$ are evaluated as
\begin{equation}
\label{eq:mu_defn}
I_\text{abs}(\omega) = \mathcal{C}\omega  \int_{-\infty}^{\infty} \left\langle\chi(0) | \chi(t)\right\rangle \exp\left(i\left(\omega+E_{0}\right)t\right) {dt},
\end{equation}
where $\omega$ is the frequency, $\mathcal{C}$ is a constant, $E_{0}$ is the energy of $\chi_{0}(\bm{R})\ket{\Phi_{\text{F}}}$.

%and $\ket{\chi(t)}$ satisfies the time-dependent Schr\"{o}dinger equation $i\frac{\partial \ket{\chi(t)}}{\partial t} = \left(\hat{T} + \hat{V}_{\text{F}}\right)\ket{\chi(t)}$.Evaluating $I_{abs}(\omega)$ using Eq. \eqref{eq:mu_defn} hinges on approximating $V_{\text{F}}(\bm{R})$ and propagating the nuclear wavepacket on it.

\subsection{Variational \gls{TGWP} dynamics}
\label{sec:tgwp}

This section presents a detailed derivation of the \glspl{EOM}
obtained from the \gls{DFVP}~\cite{dirac1930note, frenkel1934wave,
  mclachlan1964variational} for the \gls{TGWP} parameters. The nuclear
wavepacket has the form
\begin{equation}
\label{eq:gaussian_ansatz}
\chi(\bm{R}|\bm{\check{\sigma}}, \bm{P}, \gamma) =
N\exp(-\bm{R}^{\intercal}\bm{\check{\sigma}}\bm{R} +
\bm{P}^{\intercal}\bm{R}+\gamma),
\end{equation}
where
\begin{equation}
\label{eq:normalization}
N =
\sqrt{\frac{\pi^{d}}{\det(\bm{\check{\sigma}_{r}})}}\exp\left({\frac{1}{4}\bm{P_r}^{\intercal}\bm{\check{\sigma}_r}^{-1}\bm{P_r}}
  + {\gamma_{r}}\right) 
\end{equation}
is the normalization constant, $\bm{\check{\sigma}}$ is a symmetric
matrix with elements $\sigma_{jk}$, $\bm{P}$ is a one-dimensional
array with elements $P_{j}$, $\gamma$ is a scalar,
$\bm{\check{\sigma}_{r}} = 2\Re(\bm{\check{\sigma}})$,
$\bm{P_{r}} = 2\Re(\bm{P})$ and $\gamma_{r} = 2\Re(\gamma)$.
$\bm{\check{\sigma}}$, $\bm{P}$ and $\gamma$ are time-dependent.

The \glspl{EOM} are derived by plugging the variation $\delta\chi$ and
time derivative $\frac{d \chi}{d t}$ into the \gls{DFVP} (see
Appendix~\ref{sec:DFVP}), thereby producing
\begin{equation}
\label{eq:DFVP-LA-g}
\bm{h^g} = i\bm{\check{M}^{g}}
\begin{bmatrix}
\frac{d \bm{{\sigma}}}{d t} \\
\frac{d \bm{P}}{d t} \\
\frac{d \gamma}{d t} \\
\end{bmatrix},
\end{equation}
where
\begin{subequations}
\begin{equation}
\bm{h^g} = 
\begin{bmatrix}
\bm{h}^{\bm{\check{\sigma}}} \\
\bm{h}^{\bm{P}} \\
{h}^{\gamma}
\end{bmatrix},
\end{equation}
\begin{equation}
\bm{\check{M}^{g}} =
\begin{bmatrix}
\bm{\check{M}}^{\bm{\check{\sigma}}\bm{\check{\sigma}}} & \bm{\check{M}}^{\bm{\check{\sigma}}\bm{P}} & \bm{\check{M}}^{\bm{\check{\sigma}}{\gamma}} \\
\bm{\check{M}}^{\bm{P}\bm{\check{\sigma}}} & \bm{\check{M}}^{\bm{P}\bm{P}} & \bm{{M}}^{\bm{P}\gamma} \\
\bm{\check{M}}^{\gamma\bm{\check{\sigma}}} & \bm{{M}}^{\gamma\bm{P}} & {{M}}^{\gamma\gamma}
\end{bmatrix},
\end{equation}
\end{subequations}
and where $\frac{d \bm{{\sigma}}}{d t}$ is a 1D array whose elements are the
time derivatives of $\bm{\check{\sigma}}$,
$\bm{h}^{\bm{\check{\sigma}}}$ is a 1D array built by reshaping the
matrix whose elements are
$\left\langle \frac{\partial\chi}{\partial\sigma_{jk}} \middle|
  \hat{H}\chi\right\rangle$. $\bm{h}^{\bm{P}}$ is a 1D array whose
elements are
$h_{j}^{\bm{P}} = \left\langle \frac{\partial\chi}{\partial P_{j}}
  \middle| \hat{H}\chi\right\rangle$ and
${h}^{\gamma} = \left\langle \frac{\partial\chi}{\partial \gamma}
  \middle| \hat{H}\chi\right\rangle$.
$\bm{\check{M}}^{\bm{\check{\sigma}}\bm{\check{\sigma}}}$ is a square
matrix built by reshaping the 4D array whose elements are
$ {{M}}_{jkln}^{\bm{\check{\sigma}}\bm{\check{\sigma}}} = \left\langle
  \frac{\partial\chi}{\partial \sigma_{jk}}\middle|
  \frac{\partial\chi}{\partial \sigma_{ln}} \right\rangle$.
$\bm{\check{M}}^{\bm{\check{\sigma}}\bm{P}}$ is a rectangular matrix
built by reshaping the 3D array whose elements are
${{M}}_{jkn}^{\bm{\check{\sigma}}\bm{P}} = \left\langle
  \frac{\partial\chi}{\partial \sigma_{jk}}\middle|
  \frac{\partial\chi}{\partial P_{n}} \right\rangle$.
$\bm{\check{M}}^{\bm{\check{\sigma}}\gamma}$ is a square matrix whose
elements are
${{M}}_{jk}^{\bm{\check{\sigma}}\gamma} = \left\langle
  \frac{\partial\chi}{\partial \sigma_{jk}}\middle|
  \frac{\partial\chi}{\partial \gamma} \right\rangle$.
$\bm{\check{M}}^{\bm{P}\bm{P}}$ is a square matrix whose elements are
${{M}}_{jk}^{\bm{P}\bm{P}} = \left\langle \frac{\partial\chi}{\partial
    P_{j}}\middle| \frac{\partial\chi}{\partial P_{k}} \right\rangle$.
$\bm{{M}}^{\bm{P}\gamma}$ is a 1D array whose elements are
${{M}}_{j}^{\bm{P}\gamma} = \left\langle \frac{\partial\chi}{\partial
    P_{j}}\middle| \frac{\partial\chi}{\partial \gamma} \right\rangle$
and
$M^{\gamma\gamma} = \left\langle \frac{\partial\chi}{\partial
    \gamma}\middle| \frac{\partial\chi}{\partial \gamma}
\right\rangle$. The remaining submatrices,
$\bm{\check{M}}^{\bm{P}\bm{\check{\sigma}}}$,
$\bm{\check{M}}^{\gamma\bm{\check{\sigma}}}$ and
$\bm{{M}}^{\bm{P}\gamma}$ are the transposes of
$\bm{\check{M}}^{\bm{\check{\sigma}}\bm{P}}$,
$\bm{\check{M}}^{\bm{\check{\sigma}}\gamma}$ and
$\bm{{M}}^{\gamma\bm{P}}$, respectively. $\frac{d \bm{{\sigma}}}{d t}$
is 1D array whose elements are built by reshaping the 2D array whose
elements are $\frac{d \sigma_{jk}}{d t}$.

Computing the elements of $\bm{h^g}$ and $\bm{\check{M}^{g}}$ requires
computing all moments up to fourth order of the nuclear density
$\rho(\bm{R}|\bm{\check{\sigma}}, \bm{P}, \gamma) =
\chi^{*}(\bm{R}|\bm{\check{\sigma}}, \bm{P},
\gamma)\chi(\bm{R}|\bm{\check{\sigma}}, \bm{P}, \gamma)$ (see
Appendix~\ref{sec:deriv-M-h} for more details). For completeness, the
derivations of the required moments are given in
Appendix~\ref{sec:moments}.

\subsection{Separating harmonic and anharmonic contributions}
\label{sec:separ-harm-anharm}

The array $\bm{h^{g}}$ can be separated into kinetic $\bm{t^{g}}$ and
a potential energy $\bm{v^{g}}$ contributions as follows,
\begin{equation}
\label{eq:T-V-sep}
\begin{split}
\begin{bmatrix}
\bm{h}^{\bm{\check{\sigma}}} \\
\bm{h}^{\bm{P}} \\
{h}^{\gamma}
\end{bmatrix}
&=
\begin{bmatrix}
\bm{t}^{\bm{\check{\sigma}}} \\
\bm{t}^{\bm{P}} \\
{t}^{\gamma}
\end{bmatrix}
+
\begin{bmatrix}
\bm{v}^{\bm{\check{\sigma}}} \\
\bm{v}^{\bm{P}} \\
{v}^{\gamma}
\end{bmatrix} \\
&=\begin{bmatrix}
\left\langle \frac{\partial \chi}{\partial \bm{\check{\sigma}}} \middle| \hat{T}\chi \right\rangle \\
\left\langle \frac{\partial \chi}{\partial \bm{P}} \middle| \hat{T}\chi \right\rangle \\
\left\langle \frac{\partial \chi}{\partial \gamma} \middle| \hat{T}\chi \right\rangle
\end{bmatrix}
+
\begin{bmatrix}
\left\langle \frac{\partial \chi}{\partial \bm{\check{\sigma}}} \middle| \hat{V}\chi \right\rangle \\
\left\langle \frac{\partial \chi}{\partial \bm{P}} \middle| \hat{V}\chi \right\rangle \\
\left\langle \frac{\partial \chi}{\partial \gamma} \middle| \hat{V}\chi \right\rangle
\end{bmatrix},
\end{split} 
\end{equation}
where $\left\langle \bm{R} \middle| \hat{T} \chi \right\rangle = -\frac{1}{2} \sum_{j} \frac{\partial^{2}\chi(\bm{R})}{\partial R_{j}^2}$. 

The potential energy can then be further separated into two parts,
$V(\bm{R}) = V_{h}(\bm{R}) + V_{a}(\bm{R})$ where the harmonic part
$V_{h}(\bm{R})$ approximates the potential up to second order, while
the anharmonic part $V_{a}(\bm{R})$ contains all higher order terms.
Electronic structure packages can typically be utilized to obtain
$V_{h}(\bm{R})$ in Hessian calculations. When expanding from the
\gls{PES} minimum,
$V_{h}(\bm{R}) = \frac{1}{2}\sum_{j}\omega_{j}^{2}R_{j}^{2}$ where
$\omega_{j}$ is the harmonic frequency of the $j^{\text{th}}$ normal
mode and $R_{j}$ is the projection of the mass-weighted nuclear displacement along the $j^{\text{th}}$ normal
mode. When this separation is considered, Eq.~\eqref{eq:T-V-sep} becomes
\begin{equation}
\label{eq:sep-vh-va}
\begin{split}
\begin{bmatrix}
\bm{h}^{\bm{\check{\sigma}}} \\
\bm{h}^{\bm{P}} \\
{h}^{\gamma}
\end{bmatrix}
&=\begin{bmatrix}
\left\langle \frac{\partial \chi}{\partial \bm{\check{\sigma}}} \middle| \hat{T}\chi \right\rangle \\
\left\langle \frac{\partial \chi}{\partial \bm{P}} \middle| \hat{T}\chi \right\rangle \\
\left\langle \frac{\partial \chi}{\partial \gamma} \middle| \hat{T}\chi \right\rangle
\end{bmatrix}
+
\begin{bmatrix}
\left\langle \frac{\partial \chi}{\partial \bm{\check{\sigma}}} \middle| \hat{V}_{h}\chi \right\rangle \\
\left\langle \frac{\partial \chi}{\partial \bm{P}} \middle| \hat{V}_{h}\chi \right\rangle \\
\left\langle \frac{\partial \chi}{\partial \gamma} \middle| \hat{V}_{h}\chi \right\rangle
\end{bmatrix}
+
\begin{bmatrix}
\left\langle \frac{\partial \chi}{\partial \bm{\check{\sigma}}} \middle| \hat{V}_{a}\chi \right\rangle \\
\left\langle \frac{\partial \chi}{\partial \bm{P}} \middle| \hat{V}_{a}\chi \right\rangle \\
\left\langle \frac{\partial \chi}{\partial \gamma} \middle| \hat{V}_{a}\chi \right\rangle
\end{bmatrix}.
\end{split} 
\end{equation}

The terms
$\bm{t}^{\bm{g}} = \left\langle \frac{\partial \chi}{\partial
    \bm{\lambda}} \middle| \hat{T} \chi \right\rangle$ and
$\bm{v}_{h}^{\bm{g}} = \left\langle \frac{\partial \chi}{\partial
    \bm{\lambda}} \middle| \hat{V}_{h} \chi \right\rangle$, where
$\lambda$ stands as one of the variational parameters, are derived in
Appendix~\ref{sec:deriv-M-h} and
$\bm{v}_{a}^{\bm{g}} = \left\langle \frac{\partial \chi}{\partial
    \bm{\lambda}} \middle| \hat{V}_{a} \chi \right\rangle$ is
approximated in Sec.~\ref{sec:vp-krr-V}.

\subsection{\gls{TGWP} dynamics with $\Delta$-\gls{KRR} potentials}

\subsubsection{\Acrlong{KRR}}
\label{sec:krr}

The anharmonic contribution to the potential energy is fitted by
\gls{KRR}~\cite{vovk2013kernel} with a Gaussian kernel. A function
$f(\bm{X})$ fitted by \gls{KRR} takes on the form
\begin{equation}
  f(\bm{X}) = \sum_{s=1}^{N_{t}} c_{s}k\left(\bm{X}|\bm{X}_{s} \right),
\end{equation}
where $\bm{X}_{s}$ is the $\text{s}^{th}$ elements of the training
set, $N_{t}$ is the size of the training set, $c_{s}$ is the weight of
the $\text{s}^{th}$ elements of the training set and
$k(\bm{X}|\bm{X}_{s})$ is the kernel; it is a function that
corresponds to the similarity between $\bm{X}$ and $\bm{X}_{s}$.

The kernel must be a symmetric and positive semidefinite function.
Because we are propagating a
Gaussian wavepacket, an obvious choice is the Gaussian kernel
\begin{equation}
k_{g}(\bm{R}|\bm{R}_{s}, \Omega) = 
\exp
\left(-
\left(\bm{R}-\bm{R}_{s}\right)^{\intercal}
\bm{\check{\Omega}} 
\left(\bm{R}-\bm{R}_{s}\right)\right),
\end{equation}
where $\bm{\check{\Omega}} = \Omega\bm{\check{1}}$, $\bm{\check{1}}$
being the identity matrix. The hyperparameter $\Omega$ controls the
length scale on which the Gaussian kernel acts. Effectively, our
approach to taking anharmonicity into account is to fit
${V}_{a}(\bm{R})$ to a sum of $N_{t}$ Gaussians, each centered around
an element $\bm{R}_{s}$ of the training set
\begin{equation}
\label{eq:KRR-anh-V}
{V}_{a}(\bm{R}) 
\approx V_{a}^{KRR} (\bm{R})
= \sum_{s=1}^{N_{t}} c_{s}
\exp
\left(-
\left(\bm{R}-\bm{R}_{s}\right)^{\intercal}
\bm{\check{\Omega}} 
\left(\bm{R}-\bm{R}_{s}\right)\right).
\end{equation}
The weights $\bm{c} = [c_1, c_2, \cdots, c_{N_{t}}]^{\intercal}$ are
determined by minimizing the penalty function
\begin{equation}
\mathcal{P}(\bm{c}) = \sum_{s=1}^{N_{t}} \left(V_{a}(\bm{R}_{s}) -
  V_{a}^{KRR}(\bm{R}_{s})\right)^2 +
\chi\bm{c}^{\intercal}\bm{\check{K}}\bm{c}, 
\end{equation}
where $\bm{\check{K}}$ is a square matrix with elements
$K_{jk} = k_{g}(\bm{R}_{j},\bm{R}_{k})$ and $\chi \geq 0 $ is the
so-called ridge parameters. It is a hyperparameter that makes the
model less susceptible to overfitting. The coefficients minimizing
$\mathcal{P}(\bm{c})$ are
\begin{equation}
\label{eq:krr-coeffs}
\bm{c} = \left(\bm{\check{K}} + \chi\bm{\check{1}}\right)^{-1}\bm{V}_{a},
\end{equation}
where
$\bm{V}_{a} = \left[{V}_{a}(\bm{R}_{1}), {V}_{a}(\bm{R}_{2}), \cdots,
  {V}_{a}(\bm{R}_{N_{t}})\right]$.

\subsubsection{Variational dynamics with $\Delta$-\gls{KRR}
  potentials}
\label{sec:vp-krr-V}

Once a particular $\Delta$-\gls{KRR} model is built, the anharmonic
contribution to $\bm{v^{g}}$ can be incorporated into variational
quantum dynamics. By replacing $V_{a}(\bm{R})$ with
$V_{a}^{KRR}(\bm{R})$, we can approximate $\bm{v^{g}}_{a}$ as
\begin{equation}
\begin{split}
\bm{v^{g}}_{a} \approx \bm{v^{KRR}}_{a} 
&= \sum_{s=1}^{N_{t}} 
c_{s} 
\int_{-\infty}^{+\infty} 
\left(\frac{\partial \chi(\bm{R}|\bm{\check{\sigma}}, \bm{P}, \gamma)}{\partial \lambda}\right)^{*} \times \\
 & k_{g}(\bm{R},\bm{R}_{s})\chi(\bm{R}|\bm{\check{\sigma}}, \bm{P}, \gamma)
d\bm{R},
\end{split}
\end{equation}
which reduces to evaluating the weighted sum of moments of
the Gaussians $\mathcal{G}_{s}(\bm{R}|\bm{R}_{s},\bm{\check{\sigma}}, \bm{P}, \gamma)$, each being the products of the $\text{s}^{th}$ Gaussian $k_{g}(\bm{R}|\bm{R}_{s})$ and the nuclear density $\rho(\bm{R}|\bm{\check{\sigma}},
\bm{P}, \gamma)$
\begin{equation}
\begin{split}
\mathcal{G}_{s}(\bm{R}|\bm{R}_{s},\bm{\check{\sigma}}, \bm{P}, \gamma)  
&= k_{g}(\bm{R}|\bm{R}_{s})\rho(\bm{R}|\bm{\check{\sigma}},
\bm{P}, \gamma) \\
&= N^{2} \exp\left( -\bm{R}^{\intercal}\bm{\check{A}}\bm{R} + \bm{B}_{s}^{\intercal}\bm{R} +C_{s}\right),
\end{split}
\end{equation}
where
\begin{subequations}
\begin{equation}
\bm{\check{A}} = \bm{\check{\sigma}}_{r} + \bm{\check{\Omega}}, 
\end{equation}
\begin{equation}
\bm{B}_{s} = \bm{P}_{r} + 2\bm{\check{\Omega}}\bm{R}_{s}, 
\end{equation}
\begin{equation}
C_{s} = \gamma_{r} - \bm{R}_{s}^{\intercal}\bm{\check{\Omega}}\bm{R}_{s}.
\end{equation}
\end{subequations}

From Eq.~\eqref{eq:partial_deriv_psi}, we can see that we only needs to
consider zeroth, first and second-order moments to compute
the anharmonic contribution to $\bm{h^{g}}$
\begin{subequations}
\begin{equation}
\left\langle
\frac{\partial \chi}{\partial \sigma_{jk}}
\middle |
\hat{V}_{a}^{KRR}\chi
\right\rangle 
= -2\sum_{s=1}^{N_{t}} 
c_{s} 
\int_{-\infty}^{+\infty}
R_{j}R_{k}
\mathcal{G}_{s}(\bm{R}|\bm{R}_{s},\bm{\check{\sigma}}, \bm{P}, \gamma)  
d\bm{R}, \\
\end{equation}
\begin{equation}
\left\langle
\frac{\partial \chi}{\partial \sigma_{jj}}
\middle |
\hat{V}_{a}^{KRR}\chi
\right\rangle 
= -\sum_{s=1}^{N_{t}} 
c_{s} 
\int_{-\infty}^{+\infty}
R_{j}^{2}
\mathcal{G}_{s}(\bm{R}|\bm{R}_{s},\bm{\check{\sigma}}, \bm{P}, \gamma)  
d\bm{R}, \\
\end{equation}
\begin{equation}
\left\langle
\frac{\partial \chi}{\partial P_{j}}
\middle |
\hat{V}_{a}^{KRR}\chi
\right\rangle 
= \sum_{s=1}^{N_{t}} 
c_{s} 
\int_{-\infty}^{+\infty}
R_{j}
\mathcal{G}_{s}(\bm{R}|\bm{R}_{s},\bm{\check{\sigma}}, \bm{P}, \gamma)  
d\bm{R}, \\
\end{equation}
\begin{equation}
\left\langle
\frac{\partial \chi}{\partial \gamma}
\middle |
\hat{V}_{a}^{KRR}\chi
\right\rangle 
= \sum_{s=1}^{N_{t}} 
c_{s} 
\int_{-\infty}^{+\infty}
\mathcal{G}_{s}(\bm{R}|\bm{R}_{s},\bm{\check{\sigma}}, \bm{P}, \gamma)  
d\bm{R}. \\
\end{equation}
\end{subequations}
Equation~\eqref{eq:DFVP-LA-g} then becomes
\begin{equation}
\label{eq:DFVP-krr}
\bm{t^g} + \bm{v^g}_{h} +\bm{v}_{a}^{KRR} = i\bm{\check{M}^{g}}
\begin{bmatrix}
\frac{d \bm{{\sigma}}}{d t} \\
\frac{d \bm{P}}{d t} \\
\frac{d \gamma}{d t} \\
\end{bmatrix} 
\end{equation}
and it is the \gls{EOM} that governs the dynamics of a single
\gls{TGWP} on a potential where the anharmonic contributions have been
fitted using the \gls{KRR} in conjunction with a Gaussian kernel.

\subsubsection{Constructing the training set}
\label{sec:cmv}

The utility of a \gls{ML-PES} strongly depends on its training set. In
the context of a single \gls{GWP} propagation, the training set must
contain nuclear configurations that are sampled by the nuclear density
$\rho(\bf{R}|t)$. One way to construct it is to perform exhaustive
global sampling of the \gls{PES}, which becomes prohibitely expensive
for large molecules. Another way is to perform \gls{GWP} dynamics on the fly, and to construct a \gls{ML-PES} from local sampling. This
approach was studied and successfully implemented by
\citet{polyak2019direct}

Admitting the importance of a problem, we do not seek in this study
the most efficient way of producing training sets. Instead, we focus
on a general scheme to perform \gls{GWP} dynamics with \glspl{ML-PES}
regardless of the methodology used for generating the training set. We employ a
global sampling approach, which has the advantage of producing a
single training set that does not depend on
dynamics. This allows for the fair comparison of different \gls{ML}
schemes that utilize training sets drawn from the same pool of nuclear configurations and electronic energies.

A direct approach for fitting \glspl{PES} globally is to scan along
orthogonal coordinates that span a molecule internal configurational
space. One of the possible choices for such coordinates are
\glspl{VM}---the eigenvectors of a mass-weighted Hessian.
Unfortunately, they are frequently ill-suited for molecular dynamics
since these often involve curvilinear atomic
displacements~\cite{vaidehi2015internal, marsili2022quantum,
  vendrell2007full, schiffel2010quantum, joubert2012suitable}, such as
bond-angle, dihedral-angle and out-of-plane distortions. \Glspl{VM}
represent atomic displacements along which molecular bonds can break,
but the corresponding high-energy regions of the \gls{PES} usually
have little relevance for the molecular dynamics at low energies.

%They have the advantage of being easy to use when performing quantum dynamical simulation, given that their kinetic energy operator is separable and that they do not require the calculation of Jacobian determinants when evaluating integrals over nuclear degrees of freedom. 

 An \textit{ad hoc} approach, which we use in the
  present study, is to augment the \glspl{VM} in a way that allows
  curvilinear molecular displacements to be naturally incorporated. To
  this end we define \glspl{CVM}, which locally resemble
  \glspl{VM} but induce curvilinear displacements at large amplitudes.
  Operationally, \glspl{CVM} are determined by means of auxiliary
  classical dynamics simulations. The molecule is depicted as a set of
  particles connected together by rigid rods; each particle
  corresponds to an atom in the system and each rod corresponds to
  an atomic bond. Thus, the entire molecule is represented as a system
  of coupled rigid rotors. The classical dynamics of this system are
  initiated by setting the initial atomic velocity vectors aligned
  along individual \glspl{VM}. The rigid-rotor model of the molecule is
  set to evolve freely, thus allowing the normal modes to be
  projected onto curvilinear coordinates describing the motion of the rigid
  rods. It is important to note that the rigid-rotor model can only depict curvilinear displacements; it does not span a nuclear subspace containing stretching modes. Thus, to model rectilinear displacements
  associated with high-frequency stretching modes and to ensure that the set of all \glspl{CVM} forms a complete basis of the vibrational subspace, similar classical dynamics are also performed with only bond distances allowed to vary while
  all bond angles remained fixed. Thus-defined \glspl{CVM} were used to
  scan the \gls{PES} to produce a training set for fitting the
  anharmonic corrective potential $V_{a}^{KRR}(\bm{R})$.

  Classical simulations were performed for
  each VM, each time generating a one dimensional scan along its
  corresponding CVM. However, to obtain an accurate
  $V_{a}^{KRR}(\bm{R})$, a training set must also describe the
  coupling between the CVMs. This can be achieved in several ways. One
  way is to carry out further classical simulations but using linear
  combinations of VMs as initial velocities. This approach considers
  all pairs, triplets, quadruplets, etc., of VMs and comes with
  significant computational costs. Another approach is to randomly
  select nuclear configurations from the set of all 1D scans and to perturb them by random displacement along scaled
  vibrational modes. This second approach is a form of Monte Carlo
  sampling and it is the one used in this study. 

%For each VM, classical dynamical simulations are performed twice, where in the second time the initial atomic velocities are inverted. 

%It is worth mentioning that the coordinate system introduced here is reminiscent of that of Reimers\cite{reimers2001practical}, which has been used in many studies\cite{sanchez2006vibronic, peluso2009photoelectron, de2018theoretical, barone2021computational} to simulate vibrational spectra. However, there are notable differences. The coordinates introduced by Reimers are constructed from linear approximations of true curvilinear coordinates. From those, an orthonormal nonredundant coordinate basis is generated and projected onto the vibrational modes. Reimers's methodology is designed to allow one to work within the GHA. However, it is ill-suited for isolating rectilinear and curvilinear contributions within normal modes, since the coordinates are mixtures of all linearized redundant internal coordinates. Within our methodology, the CVMs are not based on linear approximations and are effectively true curvilinear coordinates. 

\section{Numerical simulations}

\subsection{Computational details}

\subsubsection{Electronic structure calculations}

Electronic structure \gls{DFT} calculations for \ce{NH3}
with the B3LYP functional~\cite{becke1993new, lee1988development} and the
6-311G(d) basis set~\cite{krishnan1980self} were performed using
the Firefly software package\cite{FFly}. The restricted open-shell formalism was used for \ce{NH3+}. Hessians were computed by first-order numerical differentiation of analytic gradients with atomic displacements of 0.01~$a_{0}$. 

Simulated spectra were translated horizontally to match the experimental spectral profile. We attribute this small energetic mismatch to the flaws of the DFT electronic structure calculations that are unable to predict accurately the relative energies of \ce{NH3} and \ce{NH3+}. 

\subsubsection{Wavepacket dynamics}

Each dynamical simulation was performed with a total duration of
500~fs and a timestep of 0.05~fs. To model environmental factors
existing in experiments but absent in dynamical simulations, the
autocorrelation functions $\left\langle\chi(0)|\chi(t)\right\rangle$,
[see also Eq.~\eqref{eq:mu_defn}] were multiplied by an
phenomenological exponential damping function, $f(t) = e^{-\kappa t}$
where $\kappa = 0.0165 \ \text{fs}^{-1}$.
Equations~\eqref{eq:DFVP-LA-g} and \eqref{eq:DFVP-krr} were
numerically integrated using the $5^{th}$ order Adams-Bashforth
multistep method~\cite{butcher2008numerical}.

The parameters of the wavepacket at $t=0$, $\bm{\check{\sigma}}_{0}$,
$\bm{P}_{0}$ and $\gamma_{0}$, were defined by parameterizing the
initial state using Heller's semiclassical ansatz~\cite{heller1975time}
\begin{equation}
\label{eq:Heller-GWP}
\begin{split}
& \chi_{H}
\left(
\bm{R}|\bm{\check{\sigma}}_{H}, \bm{R}_{H}, \bm{P}_{H}, \gamma_{H}
\right) \\
&=
N
\exp\left(
-\left(\bm{R}^{\intercal}-\bm{R}_{H}^{\intercal}\right)
\bm{\check{\sigma}}_{H}
\left(\bm{R}-\bm{R}_{H}\right)
+ \bm{P}_{H}^{\intercal}\left(\bm{R}-\bm{R}_{H}\right)
+\gamma_{H}
\right),
\end{split}
\end{equation}
where
\begin{subequations}
\begin{equation}
\bm{\check{\sigma}}_{H} = \frac{1}{2}\bm{\check{\Lambda}}_{\text{F}}^{\intercal}\bm{\check{\Lambda}}_{\text{I}}\bm{\check{\omega}}_{\text{I}}\bm{\check{\Lambda}}_{\text{I}}^{\intercal}\bm{\check{\Lambda}}_{\text{F}}, \\
\end{equation}
\begin{equation}
\bm{R}_{H} = \bm{\check{\Lambda}}_{\text{F}}^{\intercal}\bm{\check{M}}^{\frac{1}{2}}\left(\bm{X}_{\text{I}}-\bm{X}_{\text{F}}\right), \\
\end{equation}
\begin{equation}
\bm{P}_{H} = 0, \\
\end{equation}
\begin{equation}
\gamma_{H} = 0,
\end{equation}
\end{subequations}
$\bm{X}_{\text{I}}$ and $\bm{X}_{\text{F}}$ are the Cartesian
coordinates of the nuclear geometries of the \gls{PES} minima of the
initial and final electronic states and $\bm{\check{M}}$ is a diagonal matrix containing the atomic masses.
$\bm{\check{\mathcal{H}}}_{\text{I}}$ and
$\bm{\check{\mathcal{H}}}_{\text{F}}$ are the Hessians of the initial
and final electronic states, respectively. The columns of
$\bm{\check{\Lambda}}$ are the mass-weighted vibrational modes and
$\bm{\check{\omega}}$ is a diagonal matrix containing the vibrational
frequencies,
\begin{subequations}
\begin{equation}
\bm{\check{\mathcal{H}}}_{\text{I}(\text{F})}^{MW} = \bm{\check{M}}^{-\frac{1}{2}}\bm{\check{\mathcal{H}}}_{\text{I}(\text{F})}\bm{\check{M}}^{-\frac{1}{2}}, \end{equation}
\begin{equation}
\bm{\check{\mathcal{H}}}_{\text{I}(\text{F})}^{MW}\bm{\check{\Lambda}}_{\text{I}(\text{F})} = \bm{\check{\omega}}_{\text{I}(\text{F})}^{2}\bm{\check{\Lambda}}_{\text{I}(\text{F})}. \\
\end{equation}
\end{subequations}
The GWP parameters at $t=0$ as defined in Eq.~\eqref{eq:gaussian_ansatz} are
obtained by distributing the terms in the Eq.~\eqref{eq:Heller-GWP}
\begin{subequations}
\begin{equation}
\bm{\check{\sigma}}_{0} = \bm{\check{\sigma}}_{H}, 
\end{equation}
\begin{equation}
\bm{P}_{0} = 2\bm{R}_{H}^{\intercal}\bm{\check{\sigma}}_{H}+\bm{P}_{H}, 
\end{equation}
\begin{equation}
\gamma_{0} = -\bm{R}_{H}^{\intercal}\bm{\check{\sigma}}_{H}\bm{R}_{H} - \bm{P}_{H}^{\intercal}\bm{R}_{H} + \gamma.
\end{equation}
\end{subequations}

\subsubsection{Fitting anharmonic corrections with \gls{KRR}}

The set of nuclear geometries used to train and assess the
\gls{KRR}-fitted anharmonic corrective potential was generated by
scanning along each \gls{CVM} in both forward and backward directions.
Rigid-rotor dynamics were performed with timesteps of 0.5 arbitrary
time units (arb.t.u.) for 200 timesteps. The total duration of the
simulation was chosen so that for most 1D scans, atomic distances
would not be smaller than 0.5 $\text{\AA}$ nor bigger than 2.0
$\text{\AA}$. For each set of 1D scans, the nuclear geometries after
intervals of 5 arb.t.u. were taken to be part of the final training
set.

To account for couplings between \glspl{CVM}, 2000 geometries were
generated by sampling randomly from the set of geometries produced from the one dimensional
scans and randomly displacing their atoms along scaled
\glspl{VM}. The scaling coefficients were selected by latin hypercube
sampling~\cite{mckay2000comparison}, a sampling technique designed to
selects points on a grid so as to evenly cover the space of interest.
The scaling coefficients were sampled from a 6D grid where each
dimension spanned the interval $[-5,5]$. This range was selected as it
allowed most nuclear geometries to be deformed without compressing
atomic distances within 0.5 $\text{\AA}$ nor stretching them beyond
2.0 $\text{\AA}$. Geometries outside this range were discarded. Single
point calculations were performed for 2224 nuclear geometries. The
highest relative energy (with respect to the PES minima of
\ce{NH3+} was 344.43~kcal/mol.
 
We performed 10-fold cross-validation~\cite{an2007fast} to assess the performance of the \gls{KRR} model and to tune the hyperparameters $\Omega$ and $\chi$. All of the
geometries generated from the 1D scans along the \glspl{CVM} were
placed in the training set. The rest were shuffled randomly and
divided into 10 groups. The \gls{KRR} model was then fitted ten separate times, with each time taking one of the groups as the validation set and combining all others with the 1D scans to construct the training set. Ultimately, 10 sets of
training and validation set combinations were considered. Each
combination had a training set with 2025 elements and a validation set
with 199 elements.

The \gls{KRR} model was trained to compute the difference
$V_{a}(\bm{R}_{j})$ between the exact relative energy
$V_\text{rel}(\bm{R}_{j})$ and the relative energy given by the harmonic
approximation $V_{h}(\bm{R})$ given an arbitrary nuclear geometry
$\bm{R}_{j}$
\begin{subequations}
\begin{equation}
\label{eq:VaRj}
V_{a}(\bm{R}_{j}) = V_{rel}(\bm{R}_{j}) - V_{h}(\bm{R}_{j}),
\end{equation}
\begin{equation}
V_{rel}(\bm{R}_{j}) = V_{ex}(\bm{R}_{j}) - V_{ex}(\bm{R}_{\text{min}}), 
\end{equation}
\end{subequations}
where $V_{ex}(\bm{R}_{j})$ and $V_{ex}(\bm{R}_{\text{min}})$ are the
exact electronic energies of $\bm{R}_{j}$ and the \gls{PES} minima
$\bm{R}_{\text{min}}$, respectively. $V_{h}(\bm{R}_{j})$ is calculated
as
\begin{equation}
\begin{split}
V_{h}(\bm{R}_{j}) &=
                    \frac{1}{2}\bm{R}_{j}^{\intercal}\bm{\check{\mathcal{H}}}^{MW}\bm{R}_{j}. 
\end{split}
\end{equation}

The numerical inputs used to map nuclear geometry to anharmonic
correction were the mass-weighted displacement of nuclear geometries
from the \gls{PES} minimum projected along the vibrational modes
of the Hessian of the final electronic state
\begin{equation}
\bm{R}_{j} = \bm{\check{\Lambda}}_{\text{F}}^{\intercal}\bm{\check{M}}^{\frac{1}{2}}\left(\bm{X}_{j}-\bm{X}_{\text{F}}\right),
\end{equation}
where $\bm{X}_{j}$ are the Cartesian coordinates of the geometry of
interest.

The metric used to assess the performance of each individual ${k}^{th}$ fold was the difference between the \glspl{MAE} of the energies predicted by \gls{GHA} and \gls{KRR}
\begin{subequations}
\begin{equation}
MAE_{GHA}^{(k)} =
\frac{1}{N_{t}}\sum_{t=1}^{N_{t}}\left|V_{rel}\left(\bm{R}_t^{(k)}\right)
  - V_{h}\left(\bm{R}_t^{(k)}\right)\right|,
\end{equation}
\begin{equation}
MAE_{KRR}^{(k)} = \frac{1}{N_{t}}\sum_{t=1}^{N_{t}}\left|V_{rel}\left(\bm{R}_t^{(k)}\right) - \left(V_{h}\left(\bm{R}_t^{(k)}\right) + V_{a}^{KRR}\left(\bm{R}_t^{(k)}\right)\right)\right|,
\end{equation}
\end{subequations}
where $\bm{R}_{t}^{(k)}$ is the ${t}^{th}$ element of the ${k}^{th}$ validation set. To assess the performance of the regression
as a whole, we used the \gls{pMAER} defined as
\begin{equation}
\% MAER = \frac{MAE_{GHA} - MAE_{KRR}}{MAE_{GHA}} \times 100,
\end{equation}
where
\begin{subequations}
\begin{equation}
MAE_{GHA} = \frac{1}{K} \sum_{k = 1}^{K} MAE_{GHA}^{(k)}, 
\end{equation}
\begin{equation}
MAE_{KRR} = \frac{1}{K} \sum_{k = 1}^{K} MAE_{KRR}^{(k)}
\end{equation}
\end{subequations}
and $K$ is the number of folds, being 10 in our case. The
closer a model's \gls{pMAER} is to being 100$\%$, the more accurate it is deemed to be.

The hyperparameters $\Omega$ and $\chi$ (see Sec.~\ref{sec:krr}) were
selected from a 2D logarithmic grid search. We performed 10-fold cross-validation for each grid point and chose the hyperparameters corresponding to a $MAE_{KRR}$ of  $0.33 \pm 0.07$ kcal/mol (the second number being the
standard deviation). This was an improvement compared to the GHA where
$MAE_{GHA} = 23.14 \pm 3.57 $~kcal/mol, resulting in a \gls{pMAER} of
98.6 $\%$. Out of the 10 folds, the one selected to perform wavepacket dynamics was the one with the lowest $MAE_{KRR}$, which was 0.22~kcal/mol.

\subsection{Photoelectron spectra of ammonia (\ce{NH3})}
\label{sec:phot-spectra-ammon}

\subsubsection{The \acrlong{GHA}}
\label{sec:gha}

Our methodology assumes that the initial state is well-approximated by
a single Gaussian. However, \ce{NH3} is well known to have a
double-well \gls{PES} along the pyramidal inversion coordinate. Its
initial state cannot be well represented by a single Gaussian, which
may suggest that the vibronic spectra of ammonia lies outside of a
domain of applicability of our method. However, according to the
argument made by \citet{smith1968franck} based on the symmetry of the \gls{PES}  of \ce{NH3}, it is valid to depict the initial vibrational state as localized for our present purposes.

%A summary of this argument is given in the appendix. 

\begin{figure}[!b]
\includegraphics[scale=0.5]{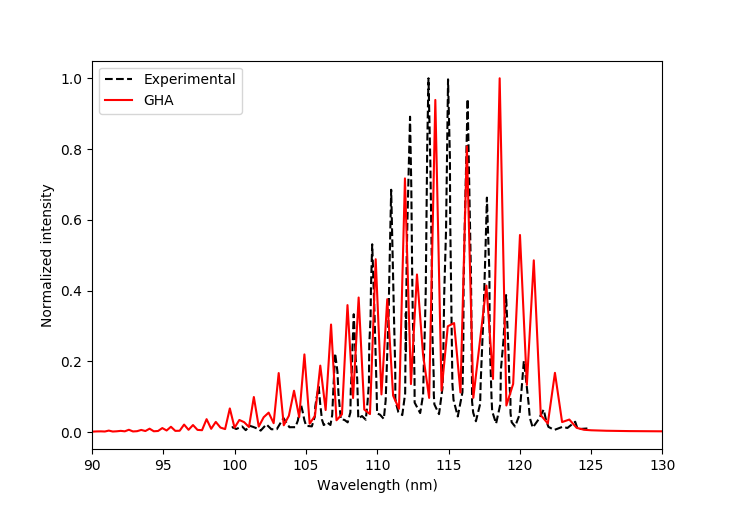}
\caption{Photoelectron spectra of ammonia. The full red line corresponds to the theoretical simulation obtained with the GHA. The dashed black line is the experimental spectrum.\cite{rabalais1973analysis}}
\label{fig:spec-nh3-exp-HA}
\end{figure}
As evident from Fig.~\ref{fig:spec-nh3-exp-HA}, the \gls{GHA} poorly
predicts the photoelectron spectra of ammonia. The simulated spectrum
is broader than experimental one, and the vibrational progressions do not
match. While the experimental spectrum has an envelope that is more or
less symmetric with an even distribution of peak intensities, the
\gls{GHA} counterpart is highly asymmetric and peaks do not seem to
follow a particular pattern. Peak spacings are also incorrect in the
\gls{GHA} spectrum---they are much closer together than in the
experiment, implying more Franck-Condon transitions than there should
be. The failures of the \gls{GHA} have been documented in many
studies~\cite{domcke1977comparison, peluso2009photoelectron,
  wehrle2015fly, begusic2022applicability}.
%Moreover, we note that our GHA spectra resembles those obtained in previous studies\cite{peluso2009photoelectron, wehrle2015fly} using different methodologies, which is indicative of the correctness of our quantum dynamical approach and its implementation.   

\subsubsection{The $\Delta$-\gls{KRR} scheme and reducing the
  dimensionality of the training set}

\begin{figure}
\includegraphics[scale=0.5]{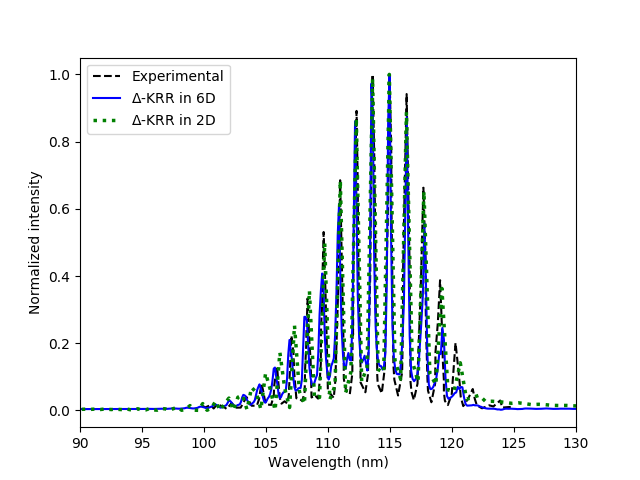}
\caption{Photoelectron spectra of ammonia. The full blue line
  corresponds to the theoretical simulation obtained with an
  anharmonic corrective $\Delta$-\gls{KRR}-fitted potential sampling
  all six \glspl{CVM}. The green dotted line corresponds to the
  theoretical simulation obtained with an anharmonic corrective
  $\Delta$-\gls{KRR}-fitted potential in which the training set
  samples the nuclear configurational space across two \glspl{CVM}.
  The experimental spectrum is taken from
  Ref.~\citenum{rabalais1973analysis}. The theoretical spectra were
  shifted to the left by 2.0 and 1.8~nm, respectively.}
\label{fig:spec-nh3-delta-KRR}
\end{figure}

As can be seen in Fig.~\ref{fig:spec-nh3-delta-KRR},
$\Delta$-\gls{KRR}-simulated spectrum agrees very well with the experimental one. The anharmonic model correctly reproduces the
spacings between the peaks. It captures the relative intensities much
better than the \gls{GHA}, albeit with some peaks having slightly
smaller amplitudes than their experimental counterpart. Overall, it is
considerable improvement over the \gls{GHA}.

The nuclear geometry of the minima of \ce{NH3} projected onto the
vibrational normal modes of \ce{NH3+} can be represented by the array
\begin{equation}
\begin{bmatrix}
48.52 & -0.10 & -0.21 & 11.71 & 0.42 & -0.73 
\end{bmatrix}.
\end{equation}
The two largest coefficients, 48.52 and 11.71, correspond to nuclear
displacements along the pyramidal inversion and symmetric \ce{N-H}
bond stretching modes, respectively. Being two orders of magnitudes
larger than the other contribution suggests that the only relevant
anharmonic displacements are those in the subspace spanned by these
two modes.

 Our sampling scheme can be readily adapted to cases
  when only a few nuclear \glspl{DOF} are strongly anharmonic. Rather
  than sampling all \glspl{CVM} for the training set evenly, it might
  be beneficial to put more weight on strongly anharmonic ones. To
  confirm this expectation, we trained another $\Delta$-\gls{KRR}
  model, this time involving the first and the fourth \glspl{CVM}, and
  the couplings between them when generating the training set. This
  reduced-dimensionality $\Delta$-\gls{KRR} model effectively samples
  nuclear configurations predominantly in a subspace spanned by the
  \glspl{CVM} corresponding to the pyramidal inversion and the symmetric bond
  stretch modes, and trains the anharmonic corrections from there.
  Nevertheless, the reduced model still maps the nuclear geometry
  represented in all six dimensions only giving more weight to the
  large-amplitude modes, such that the \gls{GWP} dynamics are still
  performed in the 6D vibrational space. 
  
% The reduction of
% dimensionality applies only to how the nuclear configurations were
% chosen in the training set.

% It can be used \textit{as is} when performing nuclear dynamics with all internal degrees of freedom. 

Figure~\ref{fig:spec-nh3-delta-KRR} displays the photoelectron
spectrum obtained using a reduced $\Delta$-\gls{KRR} model. The
methodology used to train the reduced model was the same as the
full 6D one. The reduced model demonstrates a great improvement over
the \gls{GHA}. It captures the peaks spacings almost perfectly and
their amplitudes are reproduced even better than in the full 6D model.
Both models were trained using 2000 nuclear configurations sampled
from perturbed \glspl{CVM} scans. However, while these were
scattered among six dimensions in the full $\Delta$-\gls{KRR} model,
in the reduced model they are distributed much more densely in a
lower-dimensional space. As a result, the fitting quality is higher
resulting in a better spectrum.

\subsubsection{The \gls{KRR} scheme for learning the entire \gls{PES}}

As was mentioned previously, a standard approach to construction of
\glspl{ML-PES} is to fit electronic energies rather than anharmonic
corrections. Hence, an obvious concern is whether the
$\Delta$-\gls{KRR} scheme is superior to its \gls{KRR} counterpart. To
answer this question, we simulated the photoelectron spectra using a
\gls{KRR} model trained to reproduce the electronic energies rather
than anharmonic corrections. To this end, we removed $v_{h}^{g}$ from
Eq.~\eqref{eq:sep-vh-va} and $V_{h}(\bm{R}_{j})$ from
Eq.~\eqref{eq:VaRj}.

\begin{figure}
\includegraphics[scale=0.5]{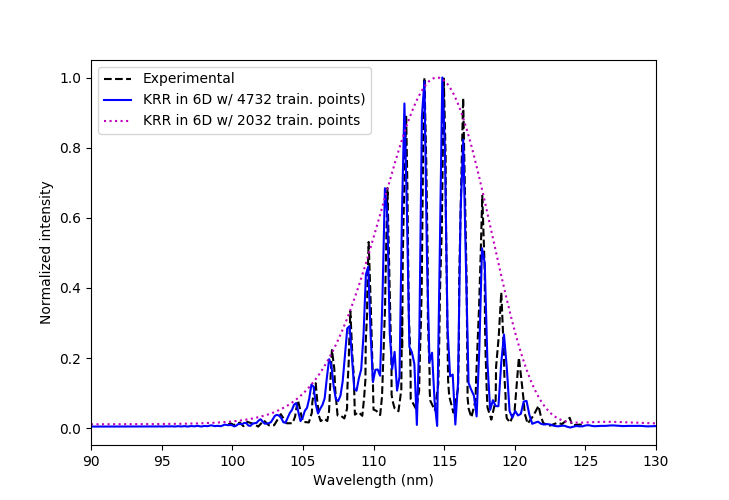}
\caption{Photoelectron spectra of ammonia. The magenta dotted line
  corresponds to the theoretical simulation obtained by fiting the
  electronic energies with 2032 training points and the blue line with
  4732 training points. The experimental spectrum is taken from
  Ref.~\citenum{rabalais1973analysis}. The theoretical spectra were
  translated to the left by 1.5 and 2.1 nm, respectively.}
\label{fig:spec-nh3-exp-small-KRR}
\end{figure}

First, we considered a training set of approximately the same size as
that for the $\Delta$-\gls{KRR} scheme, with 2023 elements. As can be
seen in Fig.~\ref{fig:spec-nh3-exp-small-KRR}, although the
corresponding simulation captured the envelope of the spectra, it
failed to demonstrate vibration progressions due to extremely poor
resolution, despite having a \gls{pMAER} of 97.4\%. Suspecting that
the \gls{KRR} scheme requires a larger training set, we retrained the
\gls{KRR} model with a training set with 4732 elements, essentially doubling
the size. As can be seen in
Fig.~\ref{fig:spec-nh3-exp-small-KRR}, the corresponding theoretical
spectrum is much closer to the experimental counterpart, representing
a great improvement over the one provided by the \gls{KRR} model
trained with a smaller training set. However, despite the increased computational efforts needed to train the \gls{KRR} model, its spectrum remains slightly inferior to the ones obtained with the $\Delta$-\gls{KRR} models.

The limitations of the \gls{KRR} model trained on the smaller,
training set of 2023 elements became more evident when the dynamics
of the center of the wavepacket were visualized. Upon leaving the
Franck-Condon region, the distances between the nitrogen and each of
the hydrogens atoms started growing steadily and without bounds
leading to complete molecular dissociation. It is clear that the
wavepacket left the properly sampled region of the \gls{PES} and
evolved on a nuclear potential that was effectively flat. By sampling
the nuclear configurational space more thoroughly with 4732 elements,
the \gls{PES} was better represented, which prevented dissociation.

Both \gls{KRR} and $\Delta$-\gls{KRR} values can be interpreted as
corrections to some reference potential. In the case of
\gls{KRR}, this reference potential is a flat 6D surface. The
\gls{KRR} model improves this flat potential by adding a Gaussian
function at the location of each element of the training set. With
insufficient sampling certain parts of the \gls{PES} remain flat.
Thus, when the \gls{GWP} reaches these regions, its dynamics
resembles thoses occuring on a flat potential. In the case of the $\Delta$-\gls{KRR} model, the reference potential is a 6D paraboloid. If the wavepacket
reaches a region that has not been sampled adequately, it will
still experience a bound potential preventing dissociative nuclear
dynamics.

\section{Conclusions}
\label{sec:conclusions}

In this study, we examined the propagation of a single thawed
\acrlong{GWP} on a harmonic \gls{PES} with \gls{KRR}-fitted anharmonic
corrections. The \gls{KRR} scheme relies upon the Gaussian kernel,
which complements nicely the wave function form, allowing for analytic
evaluation of anharmonic terms.

%To efficiently scan PESs, we also introduced curvilinear vibronic modes (CVMs). These coordinates are capable of scanning PESs along collections of curvilinear modes that are relevant for nuclear dynamics. There are constructed from a classical dynamics of coupled rigid-rotor molecular model. The geometries obtained during such propagation can then be used to construct a training set.

This type of $\Delta$-\gls{ML} approach to \gls{PES} fitting was
tested on the nuclear dynamics responsible for the photoelectron
spectrum of ammonia. We found that the quality of the
simulated spectrum is vastly superior to that of predicted with the
use of \acrlong{GHA} and compares very favorably to the experimental
results. The $\Delta$-\gls{KRR} simulations predict the correct spacings
and intensities of the vibrational progressions. 

%requiring only a smallenergy shift to match the experimental spectrum profile. We attributethis small energetic mismatch to the flaws of the \gls{DFT} electronicstructure calculations that are unable to predict the relative energies of \ce{NH3} and \ce{NH3+} species accurately.

We also compared the $\Delta$-\gls{KRR} approach to the standard one,
which consists of fiting the total electronic energies. We have shown that while it is possible to reproduce the photoelectron spectrum of ammonia, albeit not as well as the $\Delta$-\gls{KRR} approach, larger training sets are required. 

 This study also highlights the importance of selecting relevant nuclear configurations to generate small but functional
  \gls{ML} training sets. We examined an approach that considers a simplified rigid-rotor model of a
  molecule to sample ``soft'' nuclear
  \glspl{DOF} responsible for the large-amplitude motions, such as bend and
  torsion angles. By running classical dynamics simulations of this model,
  initiated along the normal modes, we identified the regions of the
  nuclear configuration space that are expected to provide large
  anharmonic corrections to the reference harmonic potential.
  Additionally, this sampling approach allowed us to construct a
  reduced-dimensionality scheme, in which only \emph{a priori}
  strongly anharmonic modes are needed to be sampled. Photoelectron
  spectra simulations of ammonia confirmed the
  viability of our approach, but its full potential is a subject of
  future studies.

\appendix

\section{The \acrlong{DFVP}}
\label{sec:DFVP}

The \acrfull{DFVP}~\cite{dirac1930note, frenkel1934wave,
  mclachlan1964variational} can be used to approximate quantum
dynamics when the wavefunction $\chi(\mathbf{R},t)$ is
constrained to a parametric form. The parametric \glspl{EOM} are found by
solving
\begin{align}
\label{eq:DFVP-general}
\mel{\delta\chi}{\hat{H}-i\frac{d}{d t}}{\chi} = 0,
\end{align}
where $\chi(\mathbf{R},t)$ is defined at all times through parameters $\bm{\lambda}(t)$ that evolve in time, $\chi(\mathbf{R},t)\rightarrow \chi(\mathbf{R}|\bm{\lambda}(t))$. In cases where there are a finite number of parameters, $\bm{\lambda}$ is a 1D array that contains the parameters that define the wavefunction. Using the chain rule, the variation $\delta\chi$ and the time-derivative $\frac{d\chi}{d t}$ can be expressed as
\begin{subequations}
\label{eq:psi_var+deriv}
\begin{equation}
\delta\chi = \sum_{j}\frac{\partial\chi}{\partial\lambda_{j}}\delta\lambda_{j}, 
\end{equation}
\begin{equation}
\frac{d\chi}{d t} = \sum_{j} \frac{\partial\chi}{\partial\lambda_{j}}\frac{d\lambda_{j}}{d t}.
\end{equation}
\end{subequations}
By plugging these expressions into Eq.~\eqref{eq:DFVP-general}, we obtain
\begin{align} 
\left\langle
{\frac{\partial \chi}{\partial \lambda_{j}}}
\middle|
{\hat{H}}
\middle|
{\chi} 
\right\rangle
=
i \sum_{jk} 
\left\langle
{\frac{\partial\chi}{\partial\lambda_j}}
\middle|
{\frac{\partial\chi}{\partial\lambda_k}}
\right\rangle\frac{d\lambda_{k}}{d t},
\end{align}
which can be rewritten in the matrix form, 
\begin{align}
\label{eq:DFVP-LA}
\bm{h} =
i\bm{\check{M}}\frac{d\bm{\lambda}}{dt},
\end{align}
where 
$h_{j} = \left\langle
\frac{\partial\chi}{\partial\lambda_{j}}
\middle|
\hat{H}
\middle|
\chi
\right\rangle$ 
and 
$M_{jk} = \left\langle
\frac{\partial\chi}{\partial\lambda_{j}}
\middle|
\frac{\partial\chi}{\partial\lambda_{k}}
\right\rangle
$. Quantum dynamics can then be approximated by evaluating  $\bm{h}$ and $\bm{\check{M}}$ and integrating Eq.~\eqref{eq:DFVP-LA}. For a thawed Gaussian ansatz, Eq.~\eqref{eq:psi_var+deriv} becomes
\begin{align}
\delta\chi &= \sum_{jk} \frac{\partial\chi}{\partial\sigma_{jk}}\delta\sigma_{jk} 
+ \sum_{j}\frac{{\partial\chi}}{\partial P_{j}}\delta P_{l}
+\frac{\partial \chi}{\partial\gamma}\delta\gamma, \\
\frac{d\chi}{d t} &= \sum_{jk} \frac{\partial\chi}{\partial\sigma_{jk}}\frac{d \sigma_{jk} }{d t}
+ \sum_{j}\frac{{\partial\chi}}{\partial P_{j}}\frac{d P_{j} }{d t}
+\frac{\partial \chi}{\partial\gamma}\frac{d \gamma}{d t},
\end{align}
and Eq.~\eqref{eq:DFVP-LA} becomes Eq.~\eqref{eq:DFVP-LA-g}.

\section{Moments of the nuclear density}
\label{sec:moments}

Computation of $\bm{h^g}$ and $\bm{\check{M}}^{\bm{g}}$ reduces to computing all moments up to fourth order of the nuclear density. Zeroth, first, second, third and fourth-order moments of the nuclear density are hereon denoted as ${m^{(0)}}$, $\bm{m}^{(1)}$,  $\bm{\check{m}}^{(2)}$,  $\bm{m}^{(3)}$ and  $\bm{m}^{(4)}$, respectively. With the exception of ${m}^{(0)}$, which is a scalar quantity, the moments are arrays whose elements are the multidimensional integrals 
\begin{subequations}
\begin{equation}
{m}^{(0)} = \int_{-\infty}^{\infty} \rho(\bm{R}) d\bm{R}, \\
\end{equation}
\begin{equation}
{m}_{j}^{(1)} = \int_{-\infty}^{\infty} R_{j} \rho(\bm{R}) d\bm{R}, \\
\end{equation}
\begin{equation}
{m}_{jk}^{(2)} = \int_{-\infty}^{\infty} R_{j}R_{k} \rho(\bm{R}) d\bm{R}, \\
\end{equation}
\begin{equation}
{m}_{jkl}^{(3)} = \int_{-\infty}^{\infty} R_{j}R_{k}R_{l} \rho(\bm{R}) d\bm{R}, 
\end{equation}
\begin{equation}
{m}_{jkln}^{(4)} = \int_{-\infty}^{\infty} R_{j}R_{k}R_{l}R_{n} \rho(\bm{R}) d\bm{R}, 
\end{equation}
\end{subequations}
where $R_j$, $R_k$, $R_l$ and $R_n$ are nuclear degrees of freedom, and $\rho(\bm{R}) = \chi^{*}(\bm{R})\chi(\bm{R})$.

The moments correspond to the derivatives of a multidimensional generating function $G(\bm{\tau}) = \int \exp(\bm{\tau}^{\intercal}\bm{R})\rho(\bm{R}) d\bm{R}$. By Taylor expanding $\exp(\bm{\tau}^{\intercal}\bm{R})$
\begin{equation}
\begin{aligned}
%G(\bm{\tau}) &= \int_{-\infty}^{+\infty} \left(1 + \sum_{k}\tau_{k}R_{k} +  \frac{1}{2!}\sum_{kl}\tau_{kl}R_{k}R_{l} + ...\right) \rho(\bm{R})  d\bm{R}
G(\bm{\tau}) &= \int_{-\infty}^{+\infty}\rho(\bm{R}) d\bm{R} 
+ \sum_{k} \tau_{k} \int_{-\infty}^{+\infty} R_{k} \rho(\bm{R}) d\bm{R} \\
&+  \frac{1}{2!}\sum_{kl}\tau_{k}\tau_{l}\int_{-\infty}^{+\infty}R_{k}R_{l} \rho(\bm{R}) d\bm{R} + \cdots  \\
 &= 
{m}^{(0)} 
+ \sum_{j} \tau_{j} {m}_{j}^{(1)} 
+ \frac{1}{2!}\sum_{jk}\tau_{j}\tau_{k}{m}_{jk}^{(2)}  
+ \cdots
\end{aligned}
\end{equation}
%+ \frac{1}{3!}\sum_{jkl}\tau_{j}\tau_{k}\tau_{l}{m}_{jkl}^{(3)} + \frac{1}{4!}\sum_{jkln}\tau_{j}\tau_{k}\tau_{l}\tau_{n}{m}_{jkln}^{(4)}
it becomes clear that the moments correspond to the partial derivatives of $G(\bm{\tau})$ with respect to $\bm{\tau}$.
Since the wavepacket is a Gaussian, $G(\bm{\tau})$ is also a Gaussian
\begin{equation}
G(\bm{\Gamma}) = \beta\exp\left(\bm{\Gamma}^{\intercal}\bm{\check{\alpha}}\bm{\Gamma} \right),
\end{equation}
where
\begin{subequations}
\begin{equation}
\beta = N^{2}\exp(\gamma_{r})\sqrt{\frac{\pi^{d}}{\det(\bm{\check{\sigma}_{r}})}},
\end{equation}
\begin{equation}
\bm{\check{\alpha}} = \frac{\bm{\check{\sigma}_{r}}^{-1}}{4},
\end{equation}
\begin{equation}
\bm{\Gamma} = \bm{p_{r}}+\bm{\tau}, 
\end{equation}
\end{subequations}

where $\bm{\check{\sigma}_{r}}$ is defined in Sec.~\ref{sec:tgwp}. The moments can then be directly calculated
\begin{subequations}
\label{eq:moments_numerical}
\begin{equation}
\begin{split}
{m}_{j}^{(1)} &= \frac{\partial G(\bm{\Gamma})}{\partial \Gamma_j} \bigg\rvert_{\bm{\Gamma} = \bm{p_r}} \\
&= G(\bm{p_r}){F_{j}^{'}}, 
\end{split}
\end{equation}
\begin{equation}
\begin{split}
{m}_{jk}^{(2)} &= \frac{\partial^{2} G(\bm{\Gamma})}{\partial \tau_j \partial \tau_k} \bigg\rvert_{\bm{\Gamma} = \bm{p_r}} \\
&=  G(\bm{p_r})\left( F_{j}^{'}F_{k}^{'} + F_{jk}^{''}\right), 
\end{split}
\end{equation}
\begin{equation}
\begin{split}
{m}_{jkl}^{(3)} &= \frac{\partial^{3} G(\bm{\Gamma})}{\partial \tau_j \partial \tau_k \partial \tau_l} \bigg\rvert_{\bm{\Gamma} = \bm{p_r}} \\
&= G(\bm{p_r}) \left( F_{j}^{'}F_{k}^{'}F_{l}^{'} + F_{jk}^{''}F_{l}^{'} + F_{jl}^{''}F_{k}^{'} + F_{kl}^{''}F_{j}^{'}\right), 
\end{split}
\end{equation}
\begin{equation}
\begin{split}
{m}_{jkln}^{(4)} &= \frac{\partial^{4} G(\bm{\Gamma})}{\partial \tau_j \partial \tau_k \partial \tau_l \partial \tau_n} \bigg\rvert_{\bm{\Gamma} = \bm{p_r}} \\
&= G(\bm{p_r}) 
\Bigl(
F_{j}^{'}F_{k}^{'}F_{l}^{'}F_{n}^{'}
+ F_{jk}^{''}F_{l}^{'}F_{n}^{'}
+ F_{jl}^{''}F_{k}^{'}F_{n}^{'} \\
& + F_{kn}^{''}F_{l}^{'}F_{n}^{'} 
+ F_{lm}^{''}F_{k}^{'}F_{n}^{'}
+ F_{kn}^{''}F_{j}^{'}F_{l}^{'} \\
& + F_{ln}^{''}F_{j}^{'}F_{k}^{'}
+ F_{jk}^{''}F_{ln}^{''}
+ F_{jl}^{''}F_{kn}^{''}
+ F_{jn}^{''}F_{kl}^{''}
\Bigr),
\end{split}
\end{equation}
\end{subequations}
where $\bm{F^{'}} = \bm{2\check{\alpha}\bm{p_{r}}}$ and $\bm{\check{F}^{''} = 2\bm{\check{\alpha}}}$.

\section{Computing}% $\bm{\check{M}^{g}}$ and $\bm{h^{g}}$}
\label{sec:deriv-M-h}

The partial derivatives of the \gls{GWP} with respect to its parameters are needed in order to compute $\bm{h^{g}}$ and $\bm{\check{M}^{g}}$. Since the \gls{DFVP} conserves the norm of the wavefunction,\cite{meyer2018mctdh} $\frac{\partial N}{\partial t}$ can be ignored and we only need to consider the derivatives of the exponential
\begin{subequations}
\label{eq:partial_deriv_psi}
\begin{equation}
\frac{\partial \chi(\bm{R}|\bm{\check{\sigma}}, \bm{P}, \gamma)}{\partial \sigma_{jj}} = -R_{j}^{2}\chi(\bm{R}|\bm{\check{\sigma}}, \bm{P}, \gamma), \\
\end{equation}
\begin{equation}
\frac{\partial \chi(\bm{R}|\bm{\check{\sigma}}, \bm{P}, \gamma)}{\partial \sigma_{jk}} = -2R_{j}R_{k}\chi(\bm{R}|\bm{\check{\sigma}}, \bm{P}, \gamma), \\
\end{equation}
\begin{equation}
\frac{\partial \chi(\bm{R}|\bm{\check{\sigma}}, \bm{P}, \gamma)}{\partial P_{j}} = R_{j}\chi(\bm{R}|\bm{\check{\sigma}}, \bm{P}, \gamma,), \\
\end{equation}
\begin{equation}
\frac{\partial \chi(\bm{R}|\bm{\check{\sigma}}, \bm{P}, \gamma)}{\partial \gamma} = \chi(\bm{R}|\bm{\check{\sigma}}, \bm{P}, \gamma).
\end{equation}
\end{subequations}
Equation~\eqref{eq:partial_deriv_psi} is obtained using the symmetric nature of $\bm{\check{\sigma}}$ that allows the exponential part of the wavepacket to be expressed as 
\begin{equation}
\label{eq:gauss_wavepacket_symmetric}
\exp\left(
-\sum_{j}\sigma_{jj}R_{j}^{2} -\sum_{j}\sum_{k>j}2\sigma_{jk}R_{j}R_{k} 
+\sum_{j}P_{j}R_{j} 
+\gamma
\right).
\end{equation}

Elements of $\bm{\check{M}^{g}}$ are readily computed from the moments of the nuclear density (see Eq.~\eqref{eq:moments_numerical})
\begin{subequations}
\begin{equation}
{M}_{jkln}^{\bm{\check{\sigma}}\bm{\check{\sigma}}} = 4{m}_{jkln}^{(4)}, 
\end{equation}
\begin{equation}
{M}_{jjln}^{\bm{\check{\sigma}}\bm{\check{\sigma}}} = 2{m}_{jjln}^{(4)}, 
\end{equation}
\begin{equation}
{M}_{jkll}^{\bm{\check{\sigma}}\bm{\check{\sigma}}} = 2{m}_{jkll}^{(4)}, 
\end{equation}
\begin{equation}
{M}_{jjll}^{\bm{\check{\sigma}}\bm{\check{\sigma}}} = {m}_{jjll}^{(4)},
\end{equation}
\begin{equation}
{M}_{jkl}^{\bm{\check{\sigma}}\bm{P}} = -2{m}_{jkl}^{(3)}, 
\end{equation}
\begin{equation}
{M}_{jjl}^{\bm{\check{\sigma}}\bm{P}} = -{m}_{jjl}^{(3)},  
\end{equation}
\begin{equation}
{M}_{jk}^{\bm{\check{\sigma}}\gamma} = -2{m}_{jk}^{(2)} , 
\end{equation}
\begin{equation}
{M}_{jj}^{\bm{\check{\sigma}}\gamma} = -{m}_{jj}^{(2)},  
\end{equation}
\begin{equation}
{M}_{jk}^{\bm{P}\bm{P}} = {m}_{jk}^{(2)}, 
\end{equation}
\begin{equation}
{M}_{j}^{\bm{P}\gamma} = {m}_{j}^{(1)}, 
\end{equation}
\begin{equation}
{M}^{\gamma\gamma} = 1. 
\end{equation}
\end{subequations}

The computation of $\bm{h^{g}}$ is more involved. The kinetic energy term 
$\left|\hat{T}\chi\right\rangle$ requires computing the action of the Laplacian on the wavepacket
\begin{widetext}
\begin{equation}
\begin{split}
\frac{\partial^{2}\chi(\bm{R})}{\partial R_{j}} 
&= 
\chi(\bm{R})
\Biggl(
4\sigma_{jj}^{2} R_{j}^{2} 
-4\sigma_{jj}P_{j}R_{j}
-2\sigma_{jj}
+P_{j}^{2} 
-\sum_{k\neq j}  4\sigma_{jk}P_{j}R_{k} \\
&+\sum_{k\neq j} 8\sigma_{jj}\sigma_{jk}R_{j}R_{k}
+\sum_{k\neq j}\sum_{l\neq j}4\sigma_{jk}\sigma_{jl}R_{k}R_{l}
\Biggl).
\end{split}
\end{equation}
The components of $\bm{t^{g}}$ are weighted sums of different moments of the nuclear density
\begin{subequations}
\begin{equation}
\begin{split}
\left\langle
\frac{\partial \chi}{\partial \sigma_{mn}}
\middle|
\hat{T}\chi
\right\rangle
&=
\sum_{j}
\Biggl(
4\sigma_{jj}^{2}m_{mnjj}^{(4)}
-4\sigma_{jj}P_{k}m_{mnj}^{(3)}
-2\sigma_{jj}m_{mn}^{(2)} 
 +P_{j}^{2}m_{mn}^{(2)}
-\sum_{l \neq j} 4\sigma_{jl}P_{j}m_{mnl}^{(3)} \\
&+\sum_{l \neq j} 8\sigma_{jj}\sigma_{jl}m_{mnlj}^{(4)}
 +\sum_{k\neq j}\sum_{l\neq j}4\sigma_{jl}\sigma_{jk}m_{mnkl}^{(4)}
\Biggr),
\end{split}
\end{equation}
\begin{equation}
\begin{split}
\left\langle
\frac{\partial \chi}{\partial \sigma_{nn}}
\middle|
\hat{T}\chi
\right\rangle
&=\frac{1}{2}
\sum_{j}
\Biggl(
4\sigma_{jj}^{2}m_{nnjj}^{(4)}
-4\sigma_{jj}P_{k}m_{nnj}^{(3)}
-2\sigma_{jj}m_{nn}^{(2)} 
 +P_{k}^{2}m_{nn}^{(2)} 
-\sum_{l \neq j} 4\sigma_{jl}P_{j}m_{nnl}^{(3)}\\
&+\sum_{l \neq j} 8\sigma_{jj}\sigma_{jl}m_{nnlj}^{(4)}
 +\sum_{k\neq j}\sum_{l\neq j}4\sigma_{jl}\sigma_{jk}m_{nnkl}^{(4)}
\Biggr),
\end{split}
\end{equation}
\begin{equation}
\begin{split}
\left\langle
\frac{\partial \chi}{\partial P_{n}}
\middle|
\hat{T}\chi
\right\rangle
&=-\frac{1}{2}
\sum_{j}
\Biggl(
4\sigma_{jj}^{2}m_{njj}^{(3)}
-4\sigma_{jj}P_{k}m_{nj}^{(2)}
-2\sigma_{jj}m_{n}^{(1)} 
 +P_{j}^{2}m_{n}^{(1)} 
-\sum_{l \neq j} 4\sigma_{jl}P_{k}m_{nl}^{(2)} \\
&+\sum_{l \neq j} 8\sigma_{jj}\sigma_{jl}m_{nlj}^{(3)}
 +\sum_{k\neq j}\sum_{l\neq j}4\sigma_{jl}\sigma_{jk}m_{nkl}^{(3)}
\Biggr),
\end{split}
\end{equation}
\begin{equation}
\begin{split}
\left\langle
\frac{\partial \chi}{\partial \gamma}
\middle|
\hat{T}\chi
\right\rangle
&=-\frac{1}{2}
\sum_{j}
\Biggl(
4\sigma_{jj}^{2}m_{jj}^{(2)}
-4\sigma_{jj}P_{k}m_{j}^{(1)}
-2\sigma_{jj} 
 +P_{j}^{2}
-\sum_{l \neq j} 4\sigma_{jl}P_{j}m_{l}^{(1)} \\
&+\sum_{l \neq j} 8\sigma_{jj}\sigma_{jl}m_{lj}^{(2)}
 +\sum_{k\neq j}\sum_{l\neq j}4\sigma_{jl}\sigma_{jk}m_{kl}^{(2)}
\Biggr),
\end{split}
\end{equation}
\end{subequations}
\end{widetext}
while the components of $\bm{v}_{h}^{g}$ are
\begin{subequations}
\begin{equation}
\left\langle
\frac{\partial \chi}{\partial \sigma_{nn}}
\middle|
\hat{V}_{h}\chi
\right\rangle 
= -\frac{1}{2}\sum_{j}\omega_{j}^{2}m_{nnjj}^{(4)},
\end{equation}
\begin{equation}
\left\langle
\frac{\partial \chi}{\partial \sigma_{mn}}
\middle|
\hat{V}_{h}\chi
\right\rangle 
= -\sum_{j}\omega_{j}^{2}m_{mnjj}^{(4)},
\end{equation}
\begin{equation}
\left\langle
\frac{\partial \chi}{\partial P_{n}}
\middle|
\hat{V}_{h}\chi
\right\rangle 
=\frac{1}{2}\sum_{j}\omega_{j}^{2}m_{njj}^{(3)},
\end{equation}
\begin{equation}
\left\langle
\frac{\partial \chi}{\partial \gamma}
\middle|
\hat{V}_{h}\chi
\right\rangle 
=\frac{1}{2}\sum_{j}\omega_{j}^{2}m_{jj}^{(2)}.
\end{equation}
\end{subequations}

\bibliography{tgwp_tdvp_abbrev}

\end{document}